\documentclass{iau} 

\usepackage{graphicx}
\usepackage{xcolor}
\usepackage[hyphens]{url}
\usepackage{hyperref}
\hypersetup{
     colorlinks  = true,
     linkcolor   = blue, 
     anchorcolor = BrickRed,
     citecolor   = blue,
     filecolor   = blue,
     menucolor   = blue,
     runcolor    = cyan,
     urlcolor    = blue
}
\usepackage[authoryear]{natbib}

\pubyear{2019}
\volume{355} 
\jname{The Realm of the Low-Surface-Brightness Universe}
\editors{D. Valls-Gabaud, I. Trujillo \& S. Okamoto, eds.}

\title[Galactic cirri in deep optical imaging] 
{Galactic cirri in deep optical imaging}

\author[Rom\'an, Trujillo and Montes]   
{Javier Rom\'an$^{1}$
 , Ignacio Trujillo$^{23}$
 \and Mireia Montes$^{4}$
 }

\affiliation{$^1$Instituto de Astrof\'isica de Andaluc\'ia (CSIC), Glorieta de la Astronom\'ia, 18008 Granada, Spain \\
$^2$Instituto de Astrof\'{\i}sica de Canarias, c/ V\'{\i}a L\'actea s/n, E-38205, La Laguna, Tenerife, Spain\\
$^3$Departamento de Astrof\'{\i}sica, Universidad de La Laguna, E-38206, La Laguna, Tenerife, Spain\\
$^4$School of Physics, University of New South Wales, Sydney, NSW 2052, Australia\\
 email: {\tt jromanastro@gmail.com} \\[\affilskip]
}

\begin{document}

\maketitle

\begin{abstract}

The presence of Galactic cirri in deep optical observations is one of the most challenging problems that the extragalactic community is already facing, and it is expected to be even tougher in the near future with the increasing depth of optical surveys. To address this problem, we have performed a photometric characterization of Galactic cirri in the IAC Stripe82 Legacy Survey using \textit{g}, \textit{r}, \textit{i} and \textit{z} SDSS bands. Through comprehensive image processing techniques we show that, in most cases, the colors of the cirri differ significantly from the colors of extragalactic sources. This could be a promising way to discern extragalactic features from dust of the Milky-Way. We suggest that the use of multi-band and ultra-deep optical photometry produced by the Large Synoptic Survey Telescope will be very efficient in detecting the Galactic dust with a higher spatial resolution than those provided by far infrared observations.

\end{abstract}

\firstsection 

\section{Introduction}
The information provided by the low surface brightness Universe is extremely relevant in revealing aspects of galaxy's hierarchical evolution. However, different systematic effects restrict the surface brightness limits of the optical data. There are currently observational and technical advances that are improving significantly the quality of deep optical data sets \citep[e.g.][]{2016ApJ...823..123T, 2019A&A...621A.133B}, however, the presence of interstellar dust from our own Galaxy reflecting the starlight is unavoidable. These clouds of dust, also called Galactic cirri, create considerable confusion with the faintest regions of galaxies or extragalactic features in the images. With the rapid increase in depth of typical optical observations, the confusion over Galactic cirri is becoming a major problem, even at high Galactic latitudes, with no solution currently proposed.

The imminent arrival in the coming years of the new generation of telescopes and especially the Large Synoptic Survey Telescope \citep[LSST; ][]{2009arXiv0912.0201L} will make the problem of confusion by cirri extremely harmful in the low surface brightness science. It is, therefore, crucial for the extragalactic community to address the problem of the cirri contamination in order to study the low surface brightness Universe. For this reason, we have carried out a comprehensive photometric analysis of the Galactic cirri, taking into account the systematic uncertainties of the extremely low surface brightness regime. Our aim is to provide the colors of the cirri in the optical wavelengths hoping to find differences with extragalactic sources, allowing discernment by using the optical data by itself. 

\section{Isolating the diffuse emission by the cirri}

In order to study the extremely faint diffuse emission by the cirri, we use data from the IAC Stripe82 Legacy Survey \citep{2016MNRAS.456.1359F, 2018RNAAS...2c.144R}. This dataset is a new reduction of the entire SDSS Stripe82 region but optimized for low surface brightness science, preserving the faintest extended features and minimizing the over-subtraction around sources. The excellent characteristic of this survey allows to investigate the cirri present in the footprint. We use the \textit{g, r, i} and \textit{z} SDSS bands, discarding the \textit{u} band for its residuals and large fluctuations of the sky background. Although the IAC Stripe82 Legacy Survey preserves the diffuse emission, the photometry of extended sources is extremely challenging to obtain. This forces us to perform specific data processing for photometric analysis of the cirri.

First, the scattered light by the PSF wings of the stars strongly contaminates the faint cirri emission. We have carried out a removal process of the scattered light by the stars. To do this, we have modeled the PSF in all bands of the IAC Stripe82 Legacy Survey. In addition, we have created a code that automatically fits all stars up to magnitude 15 in the r-band with the derived PSFs. The resulting images are almost free of contamination by scattered light. However, the faint stars, galaxies, residuals of the scattered light subtraction and artifacts still remain in the image. In order to isolate the diffuse emission of the Galactic cirri in the images, it is necessary to mask accurately all the sources. To do that, we constructed a specific algorithm that consists of six different layers of masking, focused in the detection of different sizes and deblending levels (sources over sources), each one with a different enlargement of the mask. This masking manages to cover all the external sources, exposing only the diffuse emission. Additionally, we increase the pixel size of the images to enhance the signal to noise ratio of the final image. Due to the large angular size scale of the Galactic cirri, there is a large margin to increase the pixel size without losing spatial resolution. We choose a 6 arcsec final pixel scale to obtain optimal spatial resolution and depth in the images. With this pixel size, the average surface brightness limits are $\mu_{lim} (3\sigma;6''\times6'')$ = 28.5, 28.0, 27.6 and 26.0 mag arcsec$^{-2}$ for the \textit{g}, \textit{r}, \textit{i} and \textit{z} bands respectively.

\begin{figure}
  \centering
   \includegraphics[width=1.0\textwidth]{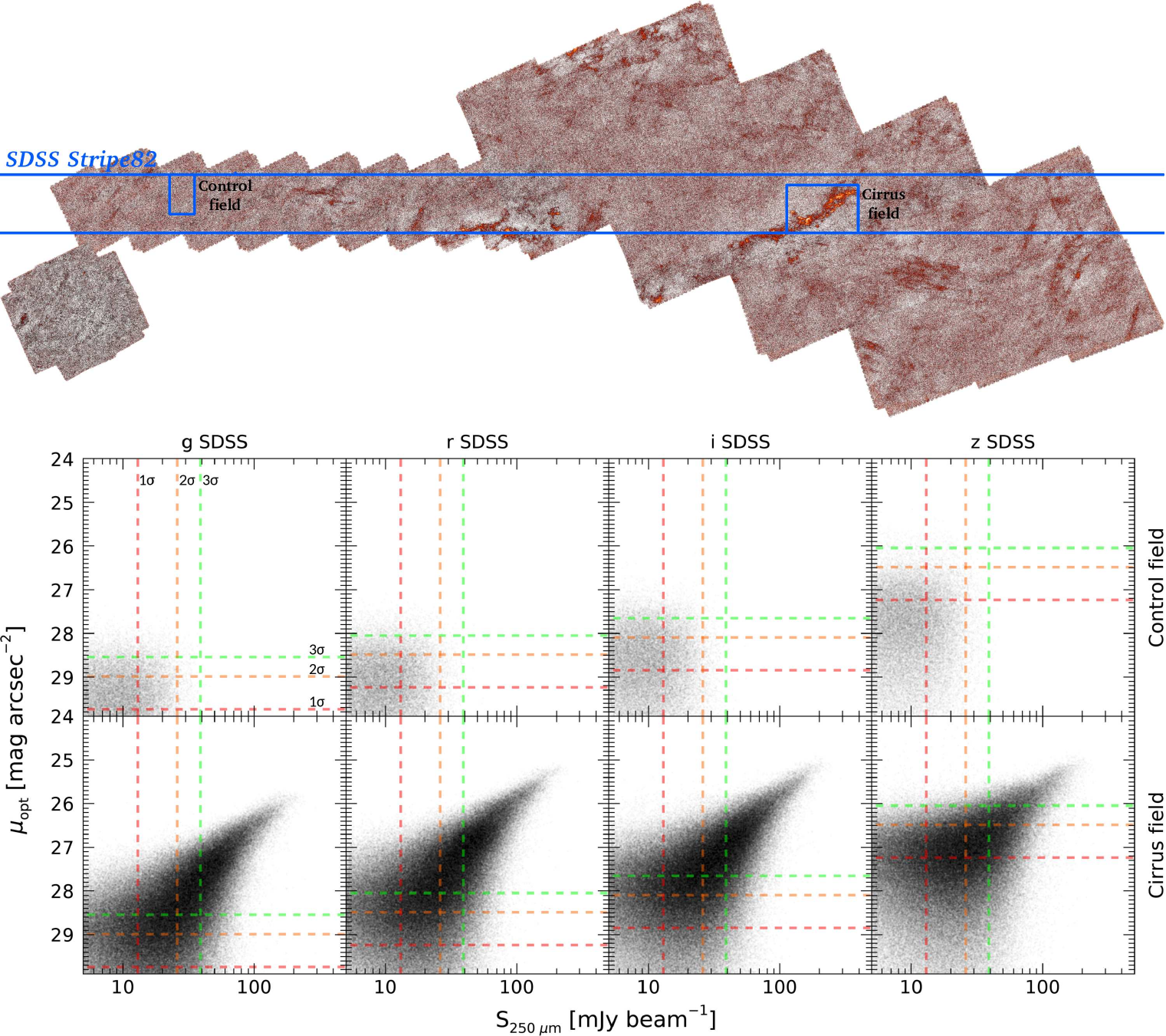}
    \caption{Correlations between optical and far IR for Galactic dust. In the upper panel it is shown the Hershel Stripe82 map with the Stripe82 region marked with blue lines. In the lower panels it is shown the correlation between the \textit{g}, \textit{r}, \textit{i} and \textit{z} SDSS optical bands and the 250 $\mu$m band for the Control field (upper row) and the Cirrus field (lower row). The detection limits are indicated with dashed lines in red, orange and green corresponding to 1$\sigma$, 2$\sigma$ and 3$\sigma$ respectively.}
   \label{fig:IR_opt}
\end{figure}

\begin{figure}
  \centering
   \includegraphics[width=0.96\textwidth]{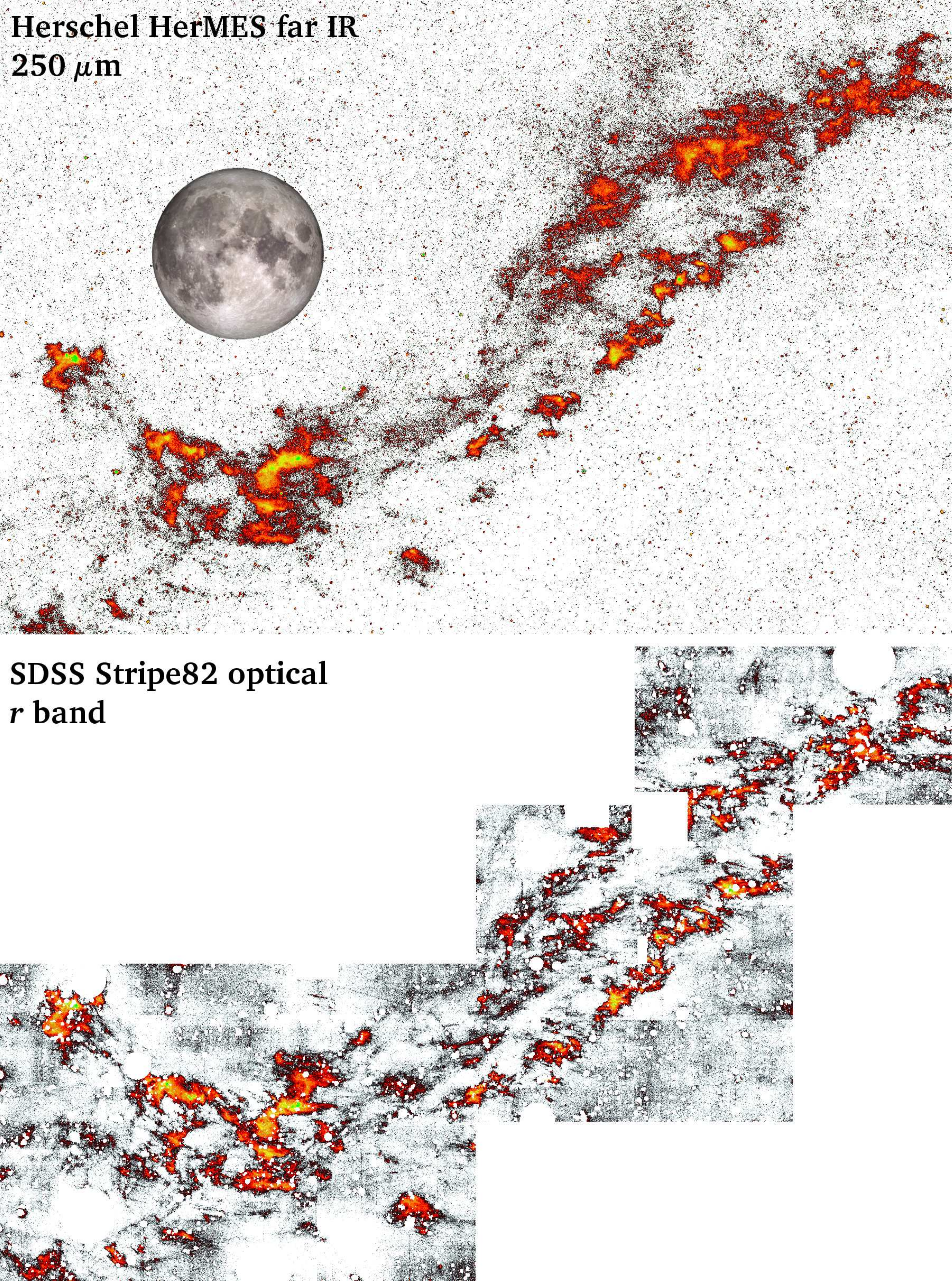}
    \caption{Comparison between the 250 $\mu$m far IR band and the \textit{r} optical band of the analyzed Cirrus field (see main text). The moon placed in the image shows an angular scale of 31 arcmin in diameter.}

   \label{fig:cirrus_filament}
\end{figure}

This data processing allows to isolate the extended diffuse emission of the optical data. However, it is necessary to investigate if this diffuse emission is due to Galactic dust entirely, or if there is another source of diffuse emission in the optical, including the residuals of the data processing. We carried out a direct photometric comparison of the diffuse emission present in the optical data of the IAC Stripe82 Legacy Survey with the diffuse emission of the far IR in the Herschel Multi-tiered Extragalactic Survey \citep[HerMES;][]{2010MNRAS.409...83L,2013ApJ...772...77V} and the Hershel Stripe82 survey \citep[HerS;][]{2014ApJS..210...22V}. In Fig. \ref{fig:IR_opt} we show the HerS-HeLMS-XMM-LSS map at 250 $\mu$m and the footprint of the SDSS Stripe82 survey indicated by horizontal blue lines. The mapped area has a relatively low presence of dust, however, some clouds are still present. We selected a region of 3 $\times$ 2 degrees in which a cirrus cloud is located, labelled in Fig. \ref{fig:IR_opt} as "Cirrus field". We did not perform any additional analysis on other regions due to the faintness of the dust present in the footprint in the far IR and optical bands, especially in the shallower \textit{z} band. We selected also an area with negligible presence of dust, with a size of 1.0 $\times$ 1.5 degrees, which we consider a control field, labelled in Fig. \ref{fig:IR_opt} as "Control field". We show the correlation between the 250 $\mu$m and optical bands in the lower panel of Fig. \ref{fig:IR_opt} for the Control field (upper row) and for the Cirrus field (lower row). We plot each pixel of the images as a dot. The Control Field shows a distribution compatible with Poisson noise in both 250 $\mu$m and optical bands, hence with absence of emission. We find no optical emission in the Control field neither by visual inspection, existing only random sky background fluctuations because of the observational drift scan mode, below the detection limits. In the case of the Cirrus field we find a strong correlation between the optical bands and the 250 $\mu$m emission. In Fig. \ref{fig:cirrus_filament} we show the Cirrus field in the far IR 250 $\mu$m band and in the optical \textit{r} band. The morphology of the diffuse emission in both spectral ranges is remarkably similar, existing a good optical counterpart to what is found in the far IR. The analysis presented here shows that every dust emission detected in the far IR has an optical counterpart. Further, we do not find any optical diffuse emission without an IR counterpart, at least in the surface brightness limits of our data. We can summarize our analysis as:

\begin{enumerate}
    \item The process of removal of the scattered light by stars and the subsequent masking of the optical images is crucial to isolate the optical diffuse emission.
    \item We do not find optical diffuse emission without a far IR counterpart (up to 28.5 mag arcsec$^{-2}$, \textit{g} band). Therefore, we can consider all the extended diffuse emission in the optical as due to Galactic dust.
    \item The optical data is efficient in detecting Galactic cirri as well as far IR data. However, the optical data can provide much better spatial resolution.
\end{enumerate}

\section{Photometric properties of the cirri}

We selected a number of regions in the IAC Stripe82 Legacy Survey for an analysis of the photometric properties of the cirri. The number of selected fields is 16, covering different areas in the survey with different levels of dust contamination. The total area analyzed is 26.5 square degrees, corresponding roughly to 10\% of the whole IAC Stripe82 Legacy Survey area. The selected fields were processed with the methodology discussed previously. To obtain the colors of the cirri, the integration of the total flux in the images is not possible due to the presence of sky background fluctuation regions. In addition, because of the extremely low surface brightness of the cirri, the effects of Poisson noise should also be taken into account. We used an approach based on the color distribution of the pixels of the images in order to obtain accurately the colors of the cirri. For that, we fitted the color distribution of the pixels within a given field with a Lorentzian plus a Gaussian function:

$$f(x) \equiv  I_{L} \frac{\frac{1}{2}\Gamma}{(x-x_{L})^{2}+(\frac{1}{2}\Gamma)^{2}} + I_{g}\exp\left(-\frac{(x-x_{g})^{2}}{2\sigma^{2}}\right) $$

Where x is the optical color and \{$I_{L}$,$\Gamma$,$x_{L}$\}, \{$I_{g}$,$\sigma$,$x_{g}$\} are the intensity, width and center of the Lorentzian and Gaussian functions respectively. The Lorentzian function describes the noise of the images while the Gaussian function provides the average color of the cirri in the selected region. Thus, the resulting $x_{g}$ is considered the average color value of the dust in the analyzed field.

\begin{figure}
  \centering
   \includegraphics[width=1.0\textwidth]{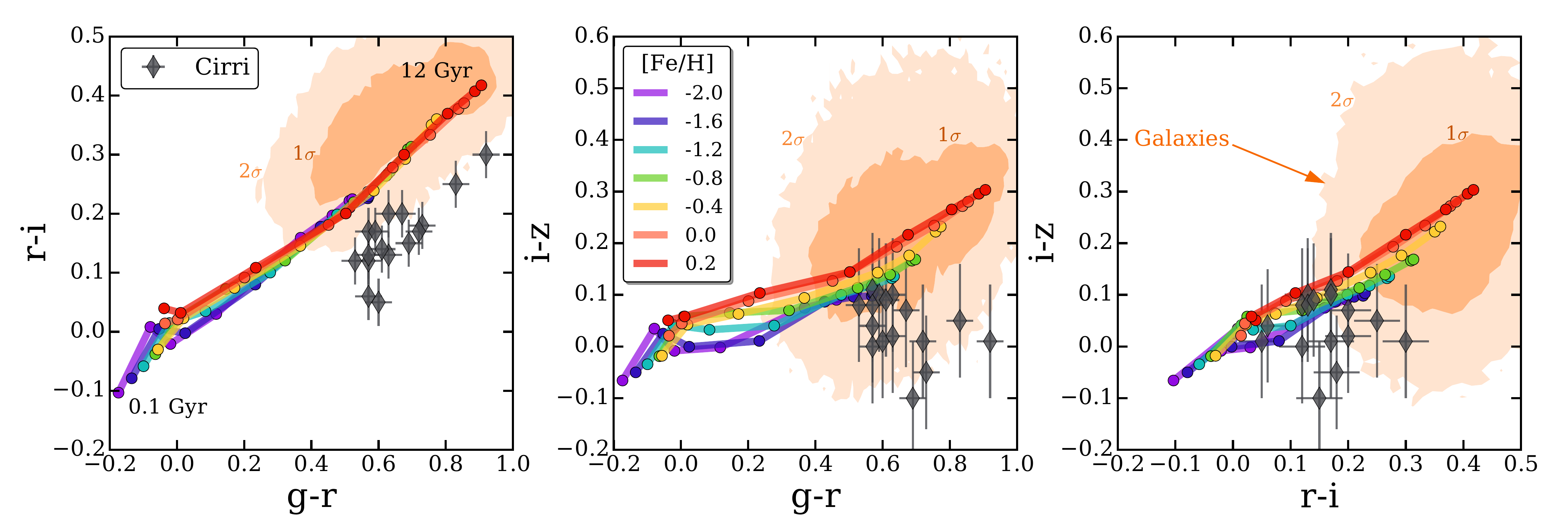}
    \caption{Comparison of Galactic dust colors with typical colors from extragalactic sources. The black diamonds show the colors of the cirri analyzed in this work with their corresponding error bars. The E-MILES single stellar population models are shown for different metallicities and ages. The colors of typical extragalatic sources are shown with contour density maps within 1$\sigma$ (dark orange contour) and 2$\sigma$ (light orange contour).}

   \label{fig:Dust_colors}
\end{figure}

In Fig.~\ref{fig:Dust_colors}, we plot the colors of the cirri in the selected fields obtained with our analysis together with the colors from the E-MILES single stellar population models \citep[][]{2016MNRAS.463.3409V} for different ages and metallicities. We include also the colors of real extragalactic sources, shown as the contour regions in the color-color maps. We plotted the 1$\sigma$ and 2$\sigma$ density contours. It is shown that, in general, the colors of the cirri have significant differences with the colors of extragalactic sources, especially in the \textit{g-r} vs \textit{r-i} map. Therefore, optical colors are a potential tool to break the confusion between Galactic dust and extragalactic sources.

\section{Conclusion}

In this work we have performed a photometric characterization of the Galactic cirri present in the Stripe82 region. Our analysis shows that the colors of the cirri differ from the typical colors of extragalactic sources. This could be very useful, as it would be possible to use optical data to discern the dust in deep optical images. The use of the LSST is of particular interest. Due to its six photometric bands (\textit{u, g, r, i, z} and \textit{y}), its great depth and large area explored (30,000 square degrees of sky), it will be possible to study the properties of Galactic dust by using this data set. The main advantage will be providing much higher spatial resolution than surveys in the far IR. However, requirements such as the characterization of the PSF to distant radii and exceptional data processing and reduction allowing information to be preserved at the lowest levels of surface brightness, are necessary. Assuming that the final combined data meet expectations, an unprecedented study of the Galactic cirri will be possible through the use of this optical survey, which will allow production of dust and extinction maps with an unprecedented quality.

This proceeding is based in the work by \cite{2020A&A...644A..42R}. Please check this reference for further details.

\begin{discussion}

\discuss{C. Mihos}{Could the consistency of colour provide an indication of the distance to cirri\,?}
\discuss{J. Rom\'an}{I know there are projects that build 3D extinction maps, but we haven't explored this issue in our work.}

\end{discussion}

\end{document}